\documentclass[12pt]{article}
\usepackage{amsfonts}
\usepackage{mathrsfs}
\usepackage{amsmath}
\usepackage{latexsym}
\usepackage{amssymb}
\usepackage{amsthm}
\usepackage{amscd}

\title{On the Hopf Algebra of Rooted Trees}

\author{ {Weicai Wu$^{a}$, Peng Wang$^{b}$,  Shouchuan Zhang$^{b}$,   Jieqiong He$^{b}$ }  \\
\small $a$. School of Mathematics,  Hunan Institute of Science and Technology, Yueyang 414006,   P.R. China\\
\small $b$. Department of Mathematics, Hunan University,  Changsha
410082, P.R. China}

\date{}
\date{}
\begin{document}
\newtheorem{Theorem}{\quad Theorem}[section]
\newtheorem{Proposition}[Theorem]{\quad Proposition}
\newtheorem{Definition}[Theorem]{\quad Definition}
\newtheorem{Corollary}[Theorem]{\quad Corollary}
\newtheorem{Lemma}[Theorem]{\quad Lemma}
\newtheorem{Example}[Theorem]{\quad Example}
\newtheorem{Remark}[Theorem]{\quad Remark}

\maketitle
\addtocounter{section}{-1}

\begin {abstract} We find a formula to compute
the number of the generators, which generate the $n$-filtered  space
of Hopf algebra of rooted trees, i.e. the number of equivalent
classes of rooted trees with weight $n$. Applying the  Hopf algebra
of rooted trees, we show that the analogue of Andruskiewitsch  and
Schneider's Conjecture is not true. The  Hopf algebra of rooted
trees and the enveloping algebra of the  Lie algebra of rooted trees
are two important examples of Hopf algebras.   We  show that they
have not any nonzero integrals. We structure their graded Drinfeld
doubles and show that they are local quasitriangular Hopf algebras.

\vskip 0.2cm

\noindent Mathematics Subject Classification: 16W30

\noindent Keywords: Rooted tree, Hopf algebra
\end {abstract}



\section{Introduction}

 In \cite{CK98},  the Hopf algebra of rooted trees ${\cal H}_R$ was
 introduced for renormalization theory.
Paper \cite {Fo02} classified the finite dimensional comodules over
${\cal H}_R$.

It is well-known that Hopf algebra ${\cal H}_R$ of rooted trees is
 the algebra of polynomials over $\mathbb{Q}$, whose indeterminate elements  are  equivalent classes
 of  rooted trees. That
 is, ${\cal H}_R = \mathbb{Q}[{\cal R}T]$ as algebras, where ${\cal
 R}T$ is the set of equivalent classes of all rooted trees (see section \ref
 {s1}).

   Therefore, it is necessary to find a formula to compute  the
number of equivalent classes of rooted trees with weight $n$.
Fortunately, we complete  this in this paper. The Hopf algebra of
rooted trees and the enveloping algebra of the Lie algebra of rooted
trees are two important examples of Hopf algebras. Primitive
elements of ${\cal H}_R$ are extraordinary. In this paper we use
this special structure to show that the analogue of Andruskiewitsch
and Schneider's Conjecture \cite [Conjecture 1.4]{AS00} is not true.
That is, there is an infinite dimensional pointed Hopf algebra,
which is not generated by its coradical and its skew-primitive
elements. That is, this conjecture is not true when the condition
which $H$ is finite dimensional is omitted. Paper \cite {ZGZ05}
pointed out that we can systematically  structure the solutions of
Yang-Baxter equations by means of local quasitriangular Hopf
algebras. In this paper we structure the graded Drinfeld double of
Hopf algebra of rooted trees and show that it is a local
quasitriangular Hopf algebra.

Note that R. Grossman and  R. G. Larson also constructed  Hopf
algebras by means of rooted trees in \cite {GL89}, but they are not
the same as the Hopf algebra of rooted trees above.

\section*{Preliminaries}
 We will use notations of \cite{CK98} and \cite {Fo02}. We
call {\it rooted tree t} a connected and simply-connected finite set
of oriented edges and vertices such that there is one distinguished
vertex with no incoming edge; this vertex is called the root of $t$.
The $weight$ of $t$ is the number of its vertices. The fertility of
a vertex $v$ of a tree $t$ is the number of edges outgoing from $v$.
A $ladder$ is a rooted tree such that every vertex has fertility
less than or equal to 1.
There is a unique ladder of weight $i$; we denote it by $l_i$.

Let $\mathbb {Z}$, $\mathbb{ Z}^+$ and $\mathbb {N}$ denote the sets
of all integers, all positive integers and all non-negative
integers, respectively.  $k=\mathbb{ Q}$ denotes rational number
field. A sub-tree $u$ of rooted tree  $t$ is said to be of the
height  $r$ if the distant from root of $t$ to root of $u$ is $r$,
i.e. there exist exactly $r$ arrows from root of $t$ to root of $u$.
Assume that $u$ and $u'$ are two sub-trees of rooted tree $t$ with
the same height. If we change the position of  $u$ and $u'$ in $t$,
we get a new rooted tree $t'$, said that $t'$ is obtained by an
elementary transformation from $t$. Two rooted trees $t$ and $t'$
are called equivalent if $t'$ can be obtained by finite elementary
transformations from $t$, written as $t \sim t'$. Note that  $t \sim
t$ for any rooted tree $t$ since we can choice $u =u'$. For example,
\begin{figure}[h] \label {f1}
\begin{picture}(400, 60)(-15, 0)

\put(130, 0){\circle*{5}} \put(130, 0){\line(1, 1){15}} \put(130,
0){\line(-1, 1){15}} \put(145, 15){\circle*{5}} \put(115,
15){\circle*{5}}

\put(145, 15){\line(0, 1){15}}

\put(145, 30){\circle*{5}}
 \put(145, 30){\line(0, 1){15}}

  \put(145,45){\circle*{5}}\put(145, 45){\line(0, 1){15}}

 \put(145,60){\circle*{5}}

\put(115, 15){\line(0, 1){15}}

\put(115, 30){\circle*{5}}
 \put(115, 30){\line(0, 1){15}}

  \put(115,45){\circle*{5}}
 \put(90, 10){$\sim$}

 \put(170, 10){$,$}

  \put(270, 10){$\sim$}

\put(50, 0){\circle*{5}} \put(50, 0){\line(1, 1){15}} \put(50,
0){\line(-1, 1){15}} \put(65, 15){\circle*{5}} \put(35,
15){\circle*{5}} \put(65, 15){\line(0, 1){15}}

\put(65, 30){\circle*{5}}
 \put(65, 30){\line(0, 1){15}}

  \put(65,45){\circle*{5}}

\put(35, 15){\line(0, 1){15}}

\put(35, 30){\circle*{5}}
 \put(35, 30){\line(0, 1){15}}

  \put(35,45){\circle*{5}}

\put(65, 15){\line(0, 1){15}}

\put(65, 30){\circle*{5}}
 \put(1465, 30){\line(0, 1){15}}

  \put(65,45){\circle*{5}}

\put(35, 15){\line(0, 1){15}}

\put(35, 30){\circle*{5}}
 \put(35, 30){\line(0, 1){15}}

  \put(35,45){\circle*{5}}

 \put(35, 60){\circle*{5}}
 \put(35, 45){\line(0, 1){15}}

  \put(35,60){\circle*{5}}

\put(50, 0){\circle*{5}} \put(50, 0){\line(1, 1){15}} \put(50,
0){\line(-1, 1){15}} \put(65, 15){\circle*{5}} \put(35,
15){\circle*{5}} \put(65, 15){\line(0, 1){15}}

\put(145, 30){\circle*{5}}
 \put(145, 30){\line(0, 1){15}}

  \put(145,45){\circle*{5}}\put(145, 45){\line(0, 1){15}}

 \put(145,60){\circle*{5}}

\put(314, 0){\circle*{5}} \put(314, 0){\line(1, 1){15}} \put(314,
0){\line(-1, 1){15}} \put(329, 15){\circle*{5}} \put(299,
15){\circle*{5}} \put(314, 0){\line(0, 15){15}} \put(314,
15){\circle*{5}}

 \put(314,30){\circle*{5}}\put(314, 15){\line(0, 1){15}}

\put(234, 0){\circle*{5}} \put(234, 0){\line(1, 1){15}} \put(234,
0){\line(-1, 1){15}} \put(249, 15){\circle*{5}} \put(219,
15){\circle*{5}} \put(234, 0){\line(0, 15){15}} \put(234,
15){\circle*{5}}

 \put(219,30){\circle*{5}}\put(219, 15){\line(0, 1){15}}
\end{picture}
\caption{\it the first and second rooted trees are equivalent;  the
third and fourth rooted trees are equivalent.}
\end{figure}
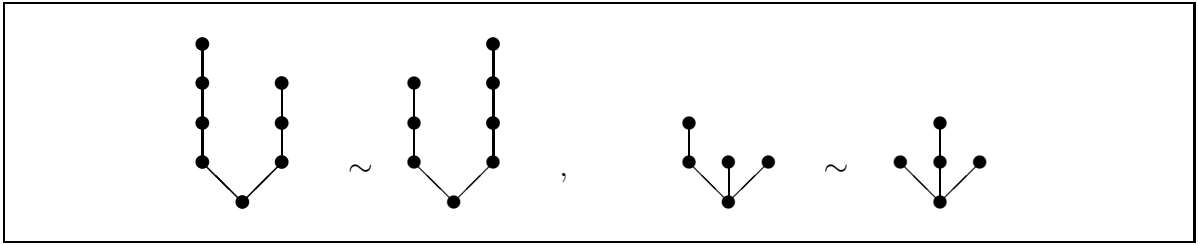

Obviously, $``\sim "$ is an equivalent relation in all rooted trees.
Let ${\cal R}T$ be the set of  equivalent classes  of all  rooted
 trees and ${\cal R}T_n$ be the set of  equivalent classes  of all  rooted
 trees with weight $n$. Define $a(n) := {\rm card }\  {\cal R}T_n$, i.e. the
 number of elements in ${\cal R}T_n$.

  $H_{\cal  R}$ denotes the algebra of polynomials
over $\mathbb{Q}$ in ${\cal R}T$. The monomials of  $H_{\cal R}$
will be called $forests$. It is often useful to think of the unit
$1$ of $H_{\cal R}$ as an empty forest. The comultiplication of
$H_{\cal R}$ was defined in \cite {CK98} and \cite [Page 90-91
]{Fo02}.  Explicitly,  we are going to give a structure of Hopf
algebra to $H_{\cal R}$. Before this, we define an {\it elementary
cut} of a rooted tree $t$ as a cut at a single chosen edge. An  {\it
admissible cut C} of a rooted tree $t$ is an assignment of
elementary cuts such that any path from any vertex of the tree has
at most one elementary cut. A cut maps a tree $t$ into a forest $t_1
\ldots t_n$. One of the $t_i$ contains the root of $t$: it will be
denoted by $R^C(t)$. The product of the others will be denoted by
$P^C(t)$. Then $\Delta$ is the morphism of algebras from $H_{\cal
R}$ into $H_{\cal R} \otimes H_{\cal R}$ such that
$$\mbox{for any rooted tree $t$, }  \Delta(t) = 1 \otimes t + t \otimes 1 + \sum_{ C \mbox{ admissible cut}}  P^C(t) \otimes R^C(t). $$
The counit is given by $\varepsilon(1)=1$, $\varepsilon(t)=0$ for any rooted tree $t$.\\
Then $H_{\cal R}$ is a Hopf algebra, with antipode given by :
$$ S(t) = \sum_{\mbox{all cuts of $t$}}(-1)^{n_C+1} P^C(t) R^C(t)$$
where $n_C$ is the number of elementary cuts in $C$. $H_{\cal R}$ is
called the Hopf algebra of rooted trees.
 $H_{ladd}$
denotes the Hopf subalgebra of $H_{\cal R}$ generated by all ladders
$l_i's.$

 ${\cal L}^1$ denotes
the Lie algebra defined in \cite {CK98} and \cite [Page 90, Section
2] {Fo02}, called the Lie algebra of  rooted tree. Explicitly, it is
the linear span of the elements $Z_t$ indexed by rooted trees. For $
t_1,\:t_2,\: t$ rooted trees, one defines $n(t_1,t_2;t)$ as the
number of elementary cuts of $t$ such that $P^C(t)=t_1$ and
$R^C(t)=t_2$. Then the Lie bracket on ${\cal L}^1$ is given by:
$$ [ Z_{t_1},Z_{t_2}]= \sum_{t} n(t_1,t_2;t) Z_t -\sum_{t} n(t_2,t_1;t) Z_t. $$
${\cal L}^1$ is graded as Lie algebra by  $degree(Z_t)=weight(t)$.
The enveloping algebra $U(\cal L^1)$ is graded as Hopf algebra with
the corresponding gradation (see \cite{CK98}).

 Let $M, N$ be two forests of ${\cal H}_R$. We define:
$$ M \top N=
\left\{
\begin{array}{cc}
\frac{1}{weight (N)} \sum \mbox{forests obtained by appending M to every node of N } & \mbox{if } N \neq 1 \\
 0  & \mbox{if } N=1.
\end{array}
\right.
$$
\noindent We can extend $.\top .$ to a bilinear map from ${\cal H}_R
\times {\cal H}_R$ into ${\cal H}_R$.

\section {The number of rooted trees}\label {s1}
 In this section we give   a formula to compute
 the number $a(n)$ of equivalent
classes of rooted trees with weight $n$.

\begin {Theorem} \label {numberoftree}
\begin {eqnarray} \label {enumberoftree} a(n+1) = \sum _ {1 \lambda _1 + 2\lambda _2 + \cdots + n\lambda
_n = n} C ^{\lambda _1}_{a(1) +\lambda _1 -1}C ^{\lambda _2}_{a(2)
+\lambda _2 -1}\cdots C ^{\lambda _n}_{a(n) +\lambda _n -1}.\end
{eqnarray}
\end {Theorem}

The proof of Theorem \ref{numberoftree} will be given in Appendix \ref {s5}.

{\bf Remark:} The indeterminate elements of algebra   ${\cal H} _R$
of polynomials are the equivalent classes of rooted trees, instead
of rooted trees. Otherwise, $B_+$ in \cite [map (45)]{CK98} is not a
map.

A rooted tree $t$ is called  an $r$-branch tree if the fertility of
every vertex of $t$ is  less than or equal to  $r$. Let $a_r(n)$
denote the number of equivalent classes of $r$-branch rooted trees
with weight $n$. We can show the following by the method similar to
the proof of Theorem \ref {numberoftree}.

\begin {Theorem} \label {2.3}
\begin {eqnarray} \label {enumberofbranchtree}a_r(n+1) &=& \sum \{ C ^{\lambda _1}_{a_r(1) +\lambda _1 -1}C
^{\lambda _2}_{a_r(2) +\lambda _2 -1}\cdots C ^{\lambda _n}_{a_r(n)
+\lambda _n -1} \ \mid \ \nonumber \\ && 1 \lambda _1 + 2\lambda _2
+ \cdots + n\lambda _n = n; \ \ {\rm card }\  \{ i \mid \ \lambda _i
\not=0\}\le r \} .\end {eqnarray}
\end {Theorem}

\section {Some properties about Hopf algebras}
\label {s3} In this section we give some properties of the  Hopf
algebra of rooted trees.

\begin {Proposition} \label {5.2}  ${\cal H}_R$, $H_{ladd}$ and $U({\cal L}^1)$ have not any
nonzero integrals.

\end {Proposition}
{\bf Proof.}  Let $l_1$ denote the rooted tree with weight
$(l_1)=1$. It is clear that the subalgebra $k [l_1]$ of ${\cal H}_R$
generated by $l_1$ is a Hopf subalgebra of ${\cal H}_R$. By  \cite
[Proposition 5.6.11] {DNR01},   $k[l_1]$ has not any nonzero
integral. Therefore, it follows from \cite [Corollary 5.3.3] {DNR01}
that ${\cal H}_R$ has not any nonzero integral. Similarly, we can
show the other. \hfill $\Box$

\begin {Proposition} \label {5.3}   $H_{ladd}$ can not be generated
by primitive elements of $H_{ladd}$ as algebras.

\end {Proposition}
{\bf Proof.}  By \cite [Proposition 9.3 and Theorem 9.5]{Fo02},
$l_2$ can not generated by set $\{P_i \mid i=1,  2,  \cdots,
\cdots\}$,  where $P_i$ is defined in \cite [Proposition 9.3]
{Fo02}. \hfill $\Box$

 N. Andruskiewitsch  and H. J. Schneider in \cite [Conjecture 1.4]{AS00} gave a conjecture: every
finite dimensional pointed Hopf algebra  can be generated by its
coradical and its skew-primitive elements. By Proposition  \ref
{5.3}, the analogue of this conjecture for an infinite dimensional
pointed Hopf algebra does not hold.

Furthermore,  $H_{ladd}$  and ${\cal H}_R$ are  not Nichols algebras
over trivial group since $(H_{ladd})_{(1)} \not= Prim (H_{ladd})$
and $({\cal H}_R)_{(1)} \not= Prim ({\cal H}_R)$. They also are not
strictly  graded  coalgebras (see \cite [P 232] {Sw69}).

A Hopf algebra $H$ is called trivial, if dim $H = 1$.

\begin {Proposition} \label {5.2} (i) If $H$ is a nontrivial pointed irreducible  Hopf algebra,    then $H$ has not any
finite dimensional nontrivial Hopf subalgebra;

(ii) ${\cal H}_R$ and $U({\cal L}^1)$ have not any finite
dimensional nontrivial Hopf subalgebra.

\end {Proposition}
{\bf Proof.} (i) Assume that $A$ is a  Hopf subalgebra of $H$ with
$A \not= H_0 = k1_{H}$. Obviously, $A_0 = k1_{H}$ and $A_1 \not=
A_0$, so there exists a nonzero primitive element $x\in A_1$. By
\cite [Lemma 3.5]{ZZC04} or \cite [Lemma 3.3]{AS98},  set  $\{x ^i
\mid i =1, 2, \cdots \}$ is linearly independent in $A$,  so $A$ is
infinite dimensional.

(ii) It immediately follows from (i).
\hfill  $\Box$

By  \cite [Theorem 3.1]{Fo02},  there is a bilinear form $<, >$ on $
{\cal U}({\cal L}^1) \times {\cal H}_R $ such that
\begin{eqnarray*}
    <1, h>&=&\varepsilon (h), \  \
< Z_t,  h>=(\frac{\partial}{\partial t}h) \mid _{t=0}, \\
 \mbox{ and }<Z_{t} Z_{t'}, h>&=&<Z_{t} \otimes Z_{t'} ,   \Delta (h)>
\end{eqnarray*} for any $h\in {\cal H}_R$, $t, t' \in {\cal R}T.$

\begin {Lemma} \label {3.1} (see \cite [Section 11.2]{Sw69})
Assume that $H= \oplus _{i=0}^\infty H_{(i)}$ is a local finite
graded Hopf algebra(i.e.  dim  $H_{(i)} < \infty$ for any $i$).
Define $H^g := \oplus _{i=0}^\infty (H_{(i)})^*$,  called graded
dual of $H$. Then $H^g$ is a graded Hopf algebra and $ H\cong (H^{g
} )^g$ as graded Hopf algebra.

\end {Lemma}
{\bf Proof.} It is clear that   $\oplus _{i=0}^\infty (H_{(i)})^*
\subseteq H^\circ.$ Now we show that $H^g= \oplus _{i=0}^\infty
(H_{(i)})^* $ is a
 graded Hopf algebra. Considering \cite [Pro.1.5.1]{Ni78},  we have to
 prove  that it is a graded bialgebra. However,  it is easy.
 Therefore, $(H_{(i)})^* = (H^g)_{(i)}$ for $i =0, 1, 2, \cdots.$

Finally,   we show that  $\sigma _H$ is a graded  Hopf algebra
isomorphism from $H$ to $H^{gg}$, where $<\sigma _H (h), f> = <f,
h>$ for any $f\in H^*, h\in H$. Indeed, we have to show  that
$\sigma _H(H) = H ^{gg}.$ For any $h \in H_{(i)}$, $f\in H_{(j)}^* =
(H^g)_{(j)}$ with $i \not= j, $ we have that $< \sigma _H(h),  f> =
<f,  h>=0, $ so $\sigma_H (h) \in (H^{gg})_{(i)}$. That implies
$\sigma _H(H) = H ^{gg}.$
\hfill  $\Box$

\begin {Corollary} \label {3.2} ${\cal H}_R^{gg }\cong {\cal H}_R$ and   $ U({\cal L}^1) ^{gg}\cong U({\cal
L}^1)$ as graded Hopf algebras.
\end {Corollary}

Recall \cite [Corollary 3.3] {Fo02},
 $ \Phi : \left \{  \begin{array}{l l}{\cal H}_R \rightarrow U({\cal L}^1)^g \\
h \mapsto < ., h>
\end{array} \right.$ and  $ \Psi : \left \{  \begin{array}{l l} U({\cal L}^1) \rightarrow ({\cal H}_R)^g \\
l \mapsto < l,  .>
\end{array} \right.$ are a  coalgebra isomorphism and
an algebra isomorphism, respectively. However, we have

\begin {Theorem} \label {3.3}
 $\Phi $ is not algebraic and $\Psi$ is not coalgebraic.
\end {Theorem}
{\bf Proof.} Let $t $ and $t'$ are two    rooted trees, which are in
different equivalent classes. It is clear that $\Phi (tt') (Z_t) =
t'$ and $<(\Phi (t)* \Phi (t')), Z_t> =0$, so $\Phi (tt') \not=
\Phi(t)* \Phi (t')$. That is, $\Phi$ is not algebraic. Similarly,
$\Psi$ is not coalgebraic. \hfill $\Box$

This implies that both $\Phi$ and $\Psi$ are not isomorphisms of
Hopf algebras.

\section {The graded Drinfeld double } \label {s4} In this section we structure  the graded
Drinfeld double of  Hopf algebra of rooted trees and show that it is
a local quasitriangular Hopf algebra.

Let  $V= \oplus _{i=0}^\infty V_{(i)}$ be a local finite graded
vector space and $V_n : = \sum _{i=0}^n V_{(i)}$ for any $n \in
\mathbb{N}$. Let $ev_{V_{(n)}} := d_{V_{(n)}}$ and $coev_{V_{(n)}}:=
b_{V_{(n)}}$ denote the evaluation and  coevaluation of $V_{(n)}$,
respectively. If we denote by $\{e^{(n)}_1 ,  e^{(n)}_2,  \cdots,
e^{(n)}_{r_n}\}$ a basis of $V_{(n)}$ and $\{f^{(n)}_1 ,  f^{(n)}_2,
\cdots, f^{(n)}_{r_n}\}$  its dual basis in  $(V_{(n)})^*$ ,  then
\begin{align}\label{coev}
b_{V_{(n)}}  := \sum _{i=1}^{r_n} (e^{(n)}_i \otimes f^{(n)}_i) \ \
\hbox { and }  b_{V_{n}} := \sum _{i=0}^n b_{V_{(i)}}
\end{align} are coevaluations of $V_{(n)}$ and $V_{n}$,
respectively,
 for any $n \in \mathbb{N}$.
Let $C_{U,V}$ denote the flip from $U \otimes V$ to $V\otimes U$ by
sending $u\otimes v$ to $v\otimes u$ for any $u\in U$, $v\in V.$

\begin {Lemma} \label {6.2}
Let  $H= \oplus _{i=0}^\infty H_{(i)}$ be a local finite graded Hopf
algebra. Under notation above, set $A= (H^g)^{cop},$ $  \tau = ev_H
C_{H, H}$, $P_n=b_{H_n} $ and $R_n = [P_n]$ $ = 1 \otimes P_n
\otimes 1 = \sum _{m=0}^n \sum _{i=1} ^{r_m} 1 \otimes e _i^{(m)}
\otimes f _i ^{(m)}\otimes 1.$ Then $(D(H),  \{R_n\})$ is a local
quasitriangular
 Hopf algebra,  where
$D(H) =A \bowtie _{\tau} H, $ called graded Drinfeld double of $H$.

\end {Lemma}
{\bf Proof.} It follows from \cite [Lemma 3.4]{ZGZ05} and \cite
[Lemma 3.6] {ZGZ05}. \hfill $\Box$

Let $\{p_i^{(n)} \mid  1 \le i \le r_n \}$ be a  basis of $(Prim
({\cal H}_R))_{(n)}$ with $n \in \mathbb{Z}^+$. By the proof of
\cite [Proposition 8.1] {Fo02}, $\{p _{i_1} ^{(j_1)}\top  p
_{i_2}^{(j_2)}\top \cdots \top p _{i_s}^{(j_s)} \mid $ $j_1 + j_2 +
\cdots+ j_s = n$, $j_1, j_2, \cdots, j_s \in \mathbb{Z}^+$;  $1 \le
i _1 \le r_{j_1}, \cdots, 1\le i_s \le r_{j_s}$; $  s\in  \mathbb
{Z}^+ \}$ is a basis of $({\cal H}_R)_{(n)}$. Let $e_{j_1, j_2,
\cdots, j_s; i_1, i_2, \cdots, i_s} ^{(n)} := p _{i_1} ^{(j_1)}\top
p _{i_2}^{(j_2)}\top \cdots \top p _{i_s}^{(j_s)}$. Therefore,
$\{e^{(n)} _{j_1, \cdots, j_s; i_1, \cdots, i_s }  \mid $ $j_1 + j_2
+ \cdots+ j_s = n$, $j_1, j_2, \cdots, j_s \in \mathbb{Z}^+$;  $1
\le i _1 \le r_{j_1}, \cdots, 1\le i_s \le r_{j_s}$; $  s\in \mathbb
{Z}^+  \}$ is a
 basis of $({\cal H}_R)_{(n)}$. Let $\{f^{(n)} _{j_1,
\cdots, j_s; i_1, \cdots, i_s }  \mid $ $j_1 + j_2 + \cdots+ j_s =
n$, $j_1, j_2, \cdots, j_s \in \mathbb{N}$;  $1 \le i _1 \le
r_{j_1}, \cdots, 1\le i_s \le r_{j_s}$; $  s\in \mathbb {Z}^+ \}$ is
its dual basis in $({\cal H}_R)^g_{(n)}$. Applying these basis and
Lemma above we have

\begin {Theorem} \label {6.3}
Let  $H= {\cal H}_R $ $= \oplus _{i=0}^\infty H_{(i)}$.  Then
$(D(H),  \{R_n\})$ is a local quasitriangular
 Hopf algebra with
 $R_n = [P_n]$ $ = 1 \otimes P_n \otimes 1
=$ $ 1\otimes 1\otimes 1\otimes 1 + \sum _{m=1}^n (\sum \{ 1 \otimes
e^{(m)} _{j_1, \cdots, j_s; i_1, \cdots, i_s }\otimes  f^{(m)}
_{j_1, \cdots, j_s; i_1, \cdots, i_s } \otimes 1 \mid $ $j_1 + j_2 +
\cdots+ j_s = m$, $j_1, j_2, \cdots, j_s \in \mathbb{N}$; $1 \le i
_1 \le r_{j_1}, \cdots, 1\le i_s \le r_{j_s}$; $  s\in \mathbb {Z}^+
\})$.
 Furthermore   $D(H)= A \bowtie _{\tau} H= A \# _\beta H$ (i.e.
(right) smash  product), i.e. $(a \bowtie h) (b \bowtie g) = \sum
_{(b)} ab_1 \otimes \beta (h, b_2)g$ and $\beta (h, a) = \sum _{(a),
(h)} < a_1, h_1> <a_2, S(h_3) >h_2 $ for any $a, b \in A$, $h, g \in
H.$
\end {Theorem}

\begin {Theorem} \label {6.4}
 Let  $H= U({\cal L}^1) = \oplus _{i=0}^\infty H_{(i)}$.
Then $(D(H),  \{R_n\})$ is a local quasitriangular
 Hopf algebra. Furthermore,    $D(H)= A \bowtie _{\tau} H= A \#  H$ (i.e.
 smash  product), i.e. $(a \bowtie h) (b \bowtie g) = \sum
_{(b)} a \alpha ( h_1, b) \otimes h_2g$ and $\alpha  (h, a) = \sum
_{(a), (h)} < a_1, h_1> <a_3, S(h_2) >a_2 $ for any $a, b \in A$,
$h, g \in H.$
\end {Theorem}

\section {Appendix}\label {s5}

{\bf Proof of Theorem \ref{numberoftree}} We first list all representatives of equivalent classes
of rooted trees with weight $n\le 6$ and check the formula  (\ref
{enumberoftree}).

\framebox(400, 50){
\begin{picture}(400, 40)(-15, 0)
\put(50, 5){\circle*{5}}\put(50, 5){\line(0,1){15}}
\put(50, 20){\circle*{5}}\put(50, 20){\line(0, 1){15}}
\put(50, 35){\circle*{5}}
\put(115,20){\circle*{5}}
\put(130, 5){\circle*{5}} \put(130, 5){\line(1, 1){15}}
\put(130,5){\line(-1, 1){15}}
\put(145, 20){\circle*{5}}
\end{picture} }

\framebox(400, 60){
\begin{picture}(400, 40)(-15, 0)
\put(35, 15){\circle*{5}}
\put(50, 0){\circle*{5}}\put(50, 0){\line(-1,1){15}}
\put(50, 0){\line(0, 1){15}}\put(50, 15){\circle*{5}}
\put(50, 0){\line(1, 1){15}}
\put(65,15){\circle*{5}}
\put(100,30){\circle*{5}}
\put(115,15){\circle*{5}} \put(115, 15){\line(-1, 1){15}}
\put(130, 0){\circle*{5}} \put(130, 0){\line(1, 1){15}}
\put(130, 0){\line(-1, 1){15}}
\put(145,15){\circle*{5}}
\put(219,30){\circle*{5}}
\put(234, 0){\circle*{5}} \put(234, 0){\line(0, 1){15}}
\put(234,15){\circle*{5}}\put(234, 15){\line(-1, 1){15}}
\put(234, 15){\line(1, 1){15}}
\put(249,30){\circle*{5}}
\put(314, 0){\circle*{5}} \put(314, 0){\line(0,1){15}}
\put(314, 15){\circle*{5}}  \put(314, 15){\line(0, 1){15}}
\put(314, 30){\circle*{5}}\put(314, 30){\line(0, 1){15}}
\put(314,45){\circle*{5}}
\end{picture} }

\framebox(400, 50){
\begin{picture}(400, 40)(-15, 0)
\put(29, 12){\circle*{5}}
\put(40, 25){\circle*{5}}
\put(50, 5){\circle*{5}} \put(50, 5){\line(-3,1){21}}
\put(50, 5){\line(-1, 2){10}}\put(50, 5){\line(1, 2){10}}
\put(50, 5){\line(3, 1){21}}
\put(60,25){\circle*{5}}
\put(71,12){\circle*{5}}
\put(100,35){\circle*{5}}
\put(115,20){\circle*{5}} \put(115, 20){\line(-1, 1){15}}
\put(130, 5){\circle*{5}} \put(130, 5){\line(1, 1){15}}
\put(130, 5){\line(-1, 1){15}}\put(130, 5){\line(0, 1){15}}
\put(130,20){\circle*{5}}
\put(145,20){\circle*{5}}
\put(204,35){\circle*{5}}
\put(219,20){\circle*{5}}\put(219, 20){\line(-1, 1){15}}
\put(234, 5){\circle*{5}}\put(234, 5){\line(-1, 1){15}}
\put(234, 5){\line(1, 1){15}}
\put(249,20){\circle*{5}} \put(249, 20){\line(1, 1){15}}
\put(264,35){\circle*{5}}
\end{picture} }

\framebox(400, 60){

\begin{picture}(400, 40)(-15, 0)
\put(35, 15){\circle*{5}}
\put(50, 0){\circle*{5}} \put(50, 0){\line(-1,1){15}}
\put(50, 0){\line(1, 1){15}}\put(50,30){\circle*{5}}
\put(65, 15){\circle*{5}}\put(65, 15){\line(1, 1){15}}
\put(65, 15){\line(-1, 1){15}}
\put(80,30){\circle*{5}}
\put(115,15){\circle*{5}}
\put(130, 0){\circle*{5}} \put(130, 0){\line(-1, 1){15}}
\put(130, 0){\line(1, 1){15}}
\put(145,15){\circle*{5}} \put(145, 15){\line(1, 1){15}}
\put(160,30){\circle*{5}}\put(160, 30){\line(1, 1){15}}
\put(175,45){\circle*{5}}
\put(219,30){\circle*{5}}
\put(234, 0){\circle*{5}} \put(234, 0){\line(0, 1){15}}
\put(234,15){\circle*{5}}\put(234, 15){\line(-1, 1){15}}
\put(234, 15){\line(1, 1){15}} \put(234, 15){\line(0, 1){15}}
\put(234,30){\circle*{5}}
\put(249,30){\circle*{5}}
\end{picture}}

\framebox(400, 80){
\begin{picture}(400, 40)(-15, 0)

\put(130, -10){\circle*{5}} \put(130, -10){\line(0, 1){15}} \put(130,
5){\circle*{5}}  \put(130, 5){\line(0, 1){15}}\put(130,
20){\circle*{5}} \put(130, 20){\line(-1, 1){15}}\put(115,
35){\circle*{5}}\put(130, 20){\line(1, 1){15}} \put(145,
35){\circle*{5}}
 \put(50, -10){\circle*{5}} \put(50, -10){\line(0,
1){15}} \put(50, 5){\circle*{5}} \put(50, 5){\line(-1, 1){15}}
\put(35, 20){\circle*{5}}\put(50, 5){\line(1, 1){15}}\put(65,
20){\circle*{5}}\put(65, 20){\line(1, 1){15}} \put(80,
35){\circle*{5}}
\put(234, -10){\circle*{5}} \put(234, -10){\line(0, 1){15}}  \put(234,
5){\circle*{5}}\put(234, 5){\line(0, 1){15}} \put(234,
20){\circle*{5}}  \put(234, 20){\line(0, 1){15}} \put(234,
35){\circle*{5}} \put(234, 35){\line(0, 1){15}} \put(234,
50){\circle*{5}}
\end{picture} }

\framebox(400, 60){

\begin{picture}(400, 40)(-15, 0)

\put(130, 0){\circle*{5}} \put(130, 0){\line(0, 1){15}}
\put(130, 15){\circle*{5}}  \put(130, 15){\line(0, 1){15}}\put(130,
30){\circle*{5}} \put(130, 0){\line(-1, 1){15}}\put(115,
15){\circle*{5}}\put(115, 15){\line(-1, 1){15}} \put(100,
30){\circle*{5}}\put(130, 0){\line(1, 1){15}}\put(145,
15){\circle*{5}}
 \put(50, 0){\circle*{5}} \put(50, 0){\line(-3,
1){21}} \put(29, 7){\circle*{5}} \put(29, 7){\line(-3, 1){21}}
\put(8, 14){\circle*{5}}\put(50, 0){\line(-1, 2){10}}\put(40,
20){\circle*{5}}\put(50, 0){\line(1, 2){10}} \put(60,
20){\circle*{5}}\put(50, 0){\line(3, 1){21}} \put(71,
7){\circle*{5}}
\put(234, 0){\circle*{5}} \put(234, 0){\line(0, 1){15}}  \put(234,
15){\circle*{5}}\put(234, 15){\line(-1, 1){15}} \put(219,
30){\circle*{5}}  \put(234, 15){\line(1, 1){15}} \put(249,
30){\circle*{5}} \put(234, 0){\line(-1, 1){15}} \put(219,
15){\circle*{5}}\put(234, 0){\line(1, 1){15}} \put(249,
15){\circle*{5}}
\put(314, 0){\circle*{5}} \put(314, 0){\line(0,
1){15}} \put(314, 15){\circle*{5}}  \put(314, 15){\line(0, 1){15}}
\put(314, 30){\circle*{5}}\put(314, 30){\line(0, 1){15}} \put(314,
45){\circle*{5}} \put(314, 0){\line(-1, 1){15}} \put(299,
15){\circle*{5}} \put(314, 0){\line(1, 1){15}} \put(329,
15){\circle*{5}}
\end{picture} }

\framebox(400, 60){

\begin{picture}(400, 40)(-15, 0)

\put(130, 0){\circle*{5}} \put(130, 0){\line(-1, 1){15}} \put(115,
15){\circle*{5}}  \put(115, 15){\line(-1, 1){15}}\put(100,
30){\circle*{5}} \put(130, 0){\line(1, 1){15}}\put(145,
15){\circle*{5}}\put(145, 15){\line(1, 1){15}} \put(160,
30){\circle*{5}}\put(160, 30){\line(1, 1){15}}\put(175,
45){\circle*{5}}
 \put(50, 0){\circle*{5}} \put(50, 0){\line(-1,
1){15}} \put(35, 15){\circle*{5}} \put(35, 15){\line(-1, 1){15}}
\put(20, 30){\circle*{5}}\put(50, 0){\line(1, 1){15}}\put(65,
15){\circle*{5}}\put(65, 15){\line(-1, 1){15}} \put(50,
30){\circle*{5}}\put(65, 15){\line(1, 1){15}} \put(80,
30){\circle*{5}}
\put(234, 0){\circle*{5}} \put(234, 0){\line(-1, 1){15}}  \put(219,
15){\circle*{5}}\put(234, 0){\line(1, 1){15}} \put(249,
15){\circle*{5}}  \put(249, 15){\line(0, 1){15}} \put(249,
30){\circle*{5}} \put(249, 15){\line(-1, 1){15}} \put(234,
30){\circle*{5}}\put(249, 15){\line(1, 1){15}} \put(264,
30){\circle*{5}}
\put(314, 0){\circle*{5}} \put(314, 0){\line(-1,
1){15}} \put(299, 15){\circle*{5}}  \put(314, 0){\line(1, 1){15}}
\put(329, 15){\circle*{5}}\put(329, 15){\line(-1, 1){15}} \put(314,
30){\circle*{5}} \put(314, 30){\line(-1, 1){15}} \put(299,
45){\circle*{5}} \put(329, 15){\line(1, 1){15}} \put(344,
30){\circle*{5}}
\end{picture} }

\framebox(400, 80){
\begin{picture}(400, 40)(-15, 0)
\put(50, -10){\circle*{5}} \put(50, -10){\line(-1,1){15}}
\put(35, 5){\circle*{5}} \put(50, -10){\line(1, 1){15}}
\put(65, 5){\circle*{5}}\put(65, 5){\line(0, 1){15}}
\put(65,20){\circle*{5}}\put(65, 20){\line(-1, 1){15}}
\put(50,35){\circle*{5}}\put(65, 20){\line(1, 1){15}}
\put(80,35){\circle*{5}}\put(130, -10){\circle*{5}}
\put(130, -10){\line(-1,1){15}} \put(115, 5){\circle*{5}}
\put(130, -10){\line(1,1){15}}\put(145, 5){\circle*{5}}
\put(145, 5){\line(1,1){15}}\put(160, 20){\circle*{5}}
\put(160, 20){\line(1, 1){15}}\put(175, 35){\circle*{5}}
\put(175, 35){\line(1, 1){15}}\put(190,50){\circle*{5}}
\put(234, -10){\circle*{5}} \put(234, -10){\line(0, 1){25}}
\put(234,15){\circle*{5}}  \put(234, -10){\line(3, 1){21}}
\put(255,-3){\circle*{5}}\put(234, -10){\line(-3, 1){21}}
\put(213,-3){\circle*{5}} \put(234, -10){\line(1, 2){10}}
\put(244,10){\circle*{5}} \put(234,-10){\line(-1, 2){10}}
\put(224,10){\circle*{5}}\put(314, -10){\circle*{5}}
\put(314, -10){\line(0,1){15}} \put(314, 5){\circle*{5}}
\put(314, 5){\line(1, 1){15}}
\put(329, 20){\circle*{5}}\put(314, 5){\line(-1, 1){15}}
\put(299,20){\circle*{5}} \put(314, 5){\line(0, 1){15}}
\put(314,20){\circle*{5}} \put(314, 20){\line(0, 1){15}}
\put(314,35){\circle*{5}}
\end{picture} }

\framebox(400, 80){
\begin{picture}(400, 40)(-15, 0)
\put(50, -10){\circle*{5}} \put(50, -10){\line(0,1){15}}
\put(50, 5){\circle*{5}} \put(50, 5){\line(-1, 1){15}}
\put(35, 20){\circle*{5}}\put(35, 20){\line(-1, 1){15}}
\put(20,35){\circle*{5}}\put(50, 5){\line(1, 1){15}}
\put(65,20){\circle*{5}}\put(65, 20){\line(1, 1){15}}
\put(80,35){\circle*{5}}
\put(130, -10){\circle*{5}} \put(130, -10){\line(0,1){15}}
\put(130, 5){\circle*{5}}  \put(130, 5){\line(-1,1){15}}
\put(115, 20){\circle*{5}} \put(130, 5){\line(1,1){15}}
\put(145, 20){\circle*{5}}\put(145, 20){\line(1, 1){15}}
\put(160, 35){\circle*{5}}\put(145, 20){\line(-1, 1){15}}
\put(130,35){\circle*{5}}\put(234, -10){\circle*{5}}
\put(234, -10){\line(0, 1){15}}  \put(234,5){\circle*{5}}
\put(234, 5){\line(-1, 1){15}} \put(219,20){\circle*{5}}
\put(234, 5){\line(1, 1){15}} \put(249,20){\circle*{5}}
\put(249, 20){\line(1, 1){15}} \put(264,35){\circle*{5}}
\put(264, 35){\line(1, 1){15}} \put(279,50){\circle*{5}}
\put(314, -10){\circle*{5}} \put(314, -10){\line(0,1){15}}
\put(314, 5){\circle*{5}}  \put(314, 5){\line(0, 1){15}}
\put(314, 20){\circle*{5}}\put(314, 20){\line(-1, 1){15}}
\put(299,35){\circle*{5}} \put(314, 20){\line(0, 1){15}}
\put(314,35){\circle*{5}} \put(314, 20){\line(1, 1){15}}
\put(329,35){\circle*{5}}
\end{picture} }

\framebox(400, 90){
\begin{picture}(400, 40)(-15, 0)
\put(35,35){\circle*{5}}
\put(50, -10){\circle*{5}} \put(50, -10){\line(0,1){15}}
\put(50, 5){\circle*{5}} \put(50, 5){\line(0, 1){15}}
\put(50, 20){\circle*{5}}\put(50, 20){\line(-1, 1){15}}
\put(50, 20){\line(1, 1){15}}
\put(65,35){\circle*{5}}\put(65, 35){\line(1, 1){15}}
\put(80,50){\circle*{5}}
\put(115,50){\circle*{5}}
\put(130, -10){\circle*{5}} \put(130, -10){\line(0,1){15}}
\put(130, 5){\circle*{5}}  \put(130, 5){\line(0,1){15}}
\put(130, 20){\circle*{5}} \put(130, 20){\line(0,1){15}}
\put(130, 35){\circle*{5}}\put(130, 35){\line(1, 1){15}}
\put(130, 35){\line(-1, 1){15}}
\put(145, 50){\circle*{5}}
\put(234, -15){\circle*{5}} \put(234, -15){\line(0, 1){15}}
\put(234,0){\circle*{5}}\put(234, 0){\line(0, 1){15}}
\put(234,15){\circle*{5}}  \put(234, 15){\line(0, 1){15}}
\put(234,30){\circle*{5}} \put(234, 30){\line(0, 1){15}}
\put(234,45){\circle*{5}}\put(234, 45){\line(0, 1){15}}
\put(234,60){\circle*{5}}
\put(293,12){\circle*{5}}
\put(304,25){\circle*{5}}
\put(314, -10){\circle*{5}} \put(314, -10){\line(0, 1){15}}
\put(314,5){\circle*{5}}\put(314, 5){\line(-3, 1){21}}
\put(314, 5){\line(-1, 2){10}} \put(314, 5){\line(1, 2){10}}
\put(314, 5){\line(3, 1){21}}
\put(324,25){\circle*{5}}
\put(335,12){\circle*{5}}
\end{picture} }

\vskip 0.4cm

\begin{tabular}{|l|l|l|l|l|l|l|l|}
  \hline
$n$& $a(n)$ &$ \lambda_{1}$ & $ \lambda_{2}$ & $ \lambda_{3}$ &$ \lambda_{4}$ &$\lambda_{5}$ \\\hline

{}& $a(3)=a(2+1)=C ^{0}_{a(1)+0 -1}C ^{1}_{a(2)+1-1}+C ^{2}_{a(1)+2 -1}$
& 0 & 1 & 0 & 0&0\\\cline{3-7}
\raisebox{1.6ex} [0pt]{n=2}&   $=1+1=2$
&2 & 0& 0 & 0& 0\\\hline
{}&  {}& 1 & 1 & 0 & 0& 0\\\cline{3-7}
  $n=3$& \raisebox{1.6ex} [0pt]{$a(4)=a(3+1)=C ^{1}_{a(1)+1 -1}C ^{1}_{a(2)+1-1}+C ^{3}_{a(1)+3 -1}$}
&0 & 0 & 1& 0& 0\\\cline{3-7}
{}& \raisebox{1.6ex} [0pt]{ $+C ^{1}_{a(3)+1 -1}=1+1+2=4$}
& 3& 0 & 0 & 0& 0\\\hline
{}&  {} & 1& 0 & 1 & 0& 0\\\cline{3-7}
  {}& $a(5)=a(4+1)=C ^{1}_{a(1)+1 -1}C ^{1}_{a(3)+1-1}$ & 2& 1& 0 & 0& 0\\\cline{3-7}
  $n=4$&  $+C ^{2}_{a(1)+2 -1}C ^{1}_{a(2)+1-1}+C ^{2}_{a(2)+2 -1}$ & 0 & 2 & 0 & 0& 0\\\cline{3-7}
  {}&  $+C ^{1}_{a(4)+1 -1}+C ^{4}_{a(1)+4 -1}=2+1+1+4+1=9$ & 0& 0 & 0& 1& 0\\\cline{3-7}
  {}&  {} & 4& 0 & 0 & 0& 0\\\hline
  {}& {} & 1& 2& 0 & 0& 0\\\cline{3-7}
  {}& $a(6)=a(5+1)=C ^{1}_{a(1)+1 -1}C ^{2}_{a(2)+2-1}$ & 2& 0& 1 & 0& 0\\\cline{3-7}
 {}& $+C ^{2}_{a(1)+2 -1}C ^{1}_{a(3)+1-1}+C ^{1}_{a(1)+1 -1}C ^{1}_{a(4)+1-1}$ & 1& 0& 0 & 1& 0\\\cline{3-7}
 $n=5$&$+C ^{1}_{a(2)+1-1}C ^{1}_{a(3)+1-1}+C ^{1}_{a(5)+1 -1}$ &0& 1& 1 & 0& 0\\\cline{3-7}
 {}& $+C ^{3}_{a(1)+3 -1}C^{1}_{a(2)+1-1}+C ^{5}_{a(1)+5 -1}$ &0& 0& 0 & 0& 1 \\\cline{3-7}
 {}& $=1+2+4+2+9+1+1=20$ &3& 1 & 0 & 0& 0 \\\cline{3-7}
 {}&  {} & 5& 0& 0 & 0& 0\\\hline

\end{tabular}
$$\hbox {Table } 1$$

For $n>6$, let  $t$ be a rooted tree with weight $n+1$. Assume that
there exists $\lambda _i$ subtrees of $t$ with height $1$ and weight
$i$ for $i =1, 2, \cdots, n$. That is, we get $\lambda _i$ subtrees
of $t$ with weight $i$, $1\le i \le n$,  after we cut the root of
rooted tree $t$. Then $1 \lambda _1 + 2\lambda _2 + \cdots +
n\lambda _n = n$. For each $i$ with $1\le i \le n$, there exist
methods of  $C^{\lambda _i}_{a(i)+\lambda _i-1}$ kinds to arrange
the some subtrees with weight $i$ to rooted tree $t$ up to
equivalent classes. Using the multiplication principle in
combination theory
 we obtain the formula. \hfill $\Box$\\

\noindent {\bf Acknowledgement }:  We thank the referees for their time and comments.

\end{document}